**Unconventional sequence of correlated Chern insulators in magic-angle twisted bilayer graphene**

Andrew T. Pierce[1*], Yonglong Xie[1,2*‡], Jeong Min Park[2*], Eslam Khalaf[1*], Seung Hwan Lee[1], Yuan Cao[2], Daniel E. Parker[1], Patrick R. Forrester[1], Shaowen Chen[1], Kenji Watanabe[3], Takashi Taniguchi[4], Ashvin Vishwanath[1], Pablo Jarillo-Herrero[2‡], Amir Yacoby[1‡]

[1]*Department of Physics, Harvard University, Cambridge, MA 02138, USA*
[2]*Department of Physics, Massachusetts Institute of Technology, Cambridge, MA 02139, USA*
[3]*Research Center for Functional Materials, National Institute for Material Science, 1-1 Namiki, Tsukuba 305-0044, Japan*
[4]*International Center for Materials Nanoarchitectonics, National Institute for Material Science, 1-1 Namiki, Tsukuba 305-0044, Japan*

*These authors contributed equally to this work.
‡Corresponding authors' emails: yxie1@g.harvard.edu, pjarillo@mit.edu, yacoby@g.harvard.edu

**The interplay between strong electron-electron interactions and band topology can lead to novel electronic states that spontaneously break symmetries. The discovery of flat bands in magic-angle twisted bilayer graphene (MATBG)[1–3] with nontrivial topology[4–7] has provided a unique platform in which to search for new symmetry-broken phases. Recent scanning tunneling microscopy[8,9] and transport experiments[10–13] have revealed a sequence of topological insulating phases in MATBG with Chern numbers $C=\pm 3, \pm 2, \pm 1$ near moiré band filling factors $\nu = \pm 1, \pm 2, \pm 3$, corresponding to a simple pattern of flavor-symmetry-breaking Chern insulators. Here, we report high-resolution local compressibility measurements of MATBG with a scanning single electron transistor that reveal a new sequence of incompressible states with unexpected Chern numbers observed down to zero magnetic field. We find that the Chern numbers for eight of the observed incompressible states are incompatible with the simple picture in which the $C=\pm 1$ bands are sequentially filled. We show that the emergence of these unusual incompressible phases can be understood as a consequence of broken translation symmetry that doubles the moiré unit cell and splits each $C=\pm 1$ band into a $C=\pm 1$ band and a $C=0$ band. Our findings**

**significantly expand the known phase diagram of MATBG, and shed light onto the origin of the close competition between different correlated phases in the system.**

Strongly-interacting electrons in the flat bands of MATBG give rise to a host of exotic correlated states including superconducting[3,14,15], correlated-insulating[2,14,15], and ferromagnetic phases[15–18]. This rich phase diagram is enabled by strong interactions[19–22] and the unique band structure of MATBG, which can be described as two fourfold degenerate sets of bands with opposite Chern numbers $C=\pm1$[7,23,24] if the $C_2T$ symmetry that protects the Dirac points is broken[25]. To date, all Chern insulator (ChI) phases reported result from successively filling the $C=\pm1$ bands that arise from the breaking of $C_2T$ symmetry. For example, hBN alignment breaks $C_2$ symmetry and gaps the Dirac points to form a set of degenerate flat bands with opposite Chern numbers in opposite valleys. Interactions further lift this degeneracy, resulting in a valley-polarized ChI with $C=1$, consistent with experiment[17]. More recent studies[8–13,26] have shown that, without hBN alignment, the combination of intrinsic electron-electron interactions and an applied perpendicular magnetic field is sufficient to isolate four degenerate bands with the same Chern number for both valleys and drive the system into a series of flavor-polarized ChI phases. In both scenarios, the observed ChIs for positive (negative) filling carry only positive (negative) Chern numbers and satisfy an even-odd symmetry where even (odd) filling factors exhibit even (odd) Chern numbers.

In this work, we report the discovery of a new sequence of incompressible states with unexpected Chern numbers in MATBG enabled by high-resolution, non-invasive local compressibility measurements using a scanning single electron transistor (SET). At zero magnetic field, we detect and extract clear thermodynamic gaps at $v=0$, +1, +2 and +3 for the first time. Applying a modest perpendicular magnetic field allows us to unravel the complex topological character of these incompressible states and establish that several ChIs are stable even at zero magnetic field, despite close competition with nearly-degenerate trivial insulators. Strikingly, we find that the incompressible states originating from positive odd (even) moiré band filling can carry even (odd) Chern numbers as well as negative Chern numbers. Similarly, for negative fillings, we observe incompressible states with Chern numbers $C=0, -1, -2$ forming near $v=-3, -2,$ and $-1$, respectively, when the perpendicular magnetic field exceeds 5 T. These observations are difficult to reconcile with theories proposed so far in which the $C=\pm1$ bands are filled sequentially, and suggest that a new mechanism that reconstructs the topology of the bands is at work. We propose that the effect of the electron-electron interaction is to favor states that double the moiré unit cell, splitting each $C=\pm1$ band into one band with $C=\pm1$ and one with $C=0$, and that this simple mechanism is capable of describing the full sequence of observed incompressible states. Our results constitute evidence for an entirely new class of interaction-driven ground states in MATBG, and demonstrate the critical role of symmetry breaking and topology in understanding the properties of the system.

The measurement protocol is depicted schematically in Fig. 1a and has been described elsewhere[27]. The hBN-encapsulated MATBG device (Fig. 1b) was fabricated using the "tear-and-stack" technique and rests on a PdAu back gate (see Methods). Global transport measurements at zero magnetic field (Fig. 1c; see Fig. 1b for circuit) reveal strong thermally-activated peaks in $R_{xx}$ between the charge neutrality point (CNP) and full-filling of the conduction moiré flat band, the hallmark of correlated-insulating states in MATBG. The pronounced activated behavior of $R_{xx}$ near the CNP in particular suggests the presence of a gap at the CNP (Extended Data Fig. 1), likely a result of alignment between the graphene sheet and the hBN substrate (see Methods and Extended Data Fig. 2 for additional evidence of alignment with hBN). Furthermore, we find a marked drop in resistance near $n = -1.97\times10^{12}$ cm$^{-2}$ (Extended Data Fig. 1), where separate transport measurements in a dilution refrigerator with a base temperature of 10 mK show that the resistance reaches zero. Finite resistance is restored by the application of a modest magnetic field or bias current (inset to Fig. 1c), indicating the unusual emergence of superconductivity in MATBG aligned with hBN. These measurements demonstrate that the device exhibits characteristic transport signatures of MATBG, to be compared with compressibility measurements discussed below.

Turning to local compressibility, our first key observation is that the system exhibits clear thermodynamic gaps at $\nu=0, +1, +2,$ and $+3$ at zero magnetic field. Fig. 1d shows the inverse compressibility d$\mu$/d$n$ measured as a function of $\nu$ and position along the white dotted line indicated in Fig. 1b. The absence of significant variation in the positions or intensities of the inverse compressibility features over a scan range of 1.2 µm demonstrates that the region under study is highly homogeneous. The spatially averaged d$\mu$/d$n$ (Fig. 1e) features two pronounced incompressible peaks at $\nu=\pm4$ associated with the energy gaps to the remote bands, the separation of which allows us to determine the local twist angle to be ~1.06° and quantify the twist angle disorder to be no greater than 0.02° over a 2.6 μm ×1.2 μm region (Extended Data Fig. 3), highlighting the exceptional quality of the device. On the hole-doped side, we find a weak "sawtooth" pattern consistent with previous studies[13,28,29], with peaks occurring approximately halfway between successive integer fillings. These peaks were interpreted as abrupt reconstructions of the Fermi surface due to interaction-driven flavor symmetry breaking. According to this picture, similar features are expected to occur on the electron-doped side. Instead, we observe sharp incompressible peaks at precisely $\nu$=+1, +2, and +3, separated by regions of negative compressibility[28,29,13], signaling the dominance of interactions and the formation of insulating states. Simultaneous DC measurements of the chemical potential exhibit step-like changes and allow us to determine the energy gaps of these insulating states (Fig. 1f). We find the size of the gaps to be on the order of 10 meV, reaching as high as 13 meV at $\nu$=+2 (see Extended Data Fig. 3), which is more than 30 times larger than the activation gap extracted from transport (see Extended Data Fig. 1). In addition, the size of the insulating gap we extract is at least a few meV larger than those observed in tunneling spectroscopy measurements[19–21]. We emphasize that our technique does not suffer from the possible soft Coulomb gap[30] that complicates the interpretation of apparent gaps in tunneling spectra, and that the tip-sample distance in our setup is approximately 100 nm as compared to a few angstroms, allowing us to avoid possible tip-induced screening effects that recent transport studies[18,31] suggest may play an important role in tunneling experiments. Importantly, we also observe a sharp incompressible peak at the CNP with a corresponding gap of 9 meV, in agreement with the expectation that the hBN substrate breaks $C_2T$ symmetry and gaps the Dirac points to form two sets of isolated, degenerate Chern bands. The presence of a thermodynamic gap at the CNP therefore provides the prerequisite splitting of the conduction and valence flat bands for the emergence of insulating states with non-zero Chern numbers at $\nu$=+1, +2, and +3. These compressibility signatures, reproduced at many different locations (Extended Data Fig. 3), constitute the first unambiguous observation of thermodynamic gaps of correlated insulating states in MATBG at zero magnetic field.

A topologically-insulating gap can be classified according to its response to perpendicular magnetic fields, which encodes the Chern number of the corresponding incompressible state. Figure 2a shows inverse compressibility as a function of moiré band filling factor ν and perpendicular magnetic fields $B$ up to 11.6 T. The data reveal a host of incompressible states that disperse linearly not only from the CNP, but also from numerous nonzero integer values of ν according to the Streda formula[32] $dn/dB = C/\phi_0$, where $C$ is the Chern number and $\phi_0$ is the magnetic flux quantum. Thus, we can identify the Chern number $C$ associated with each incompressible state by extracting its slope on the $n$ vs. $B$ plot (Fig. 2b), and characterize each state by its corresponding ($C$, $s$), where $s$ is the moiré band filling. Near the CNP, we find a series of strong incompressible peaks (black lines in Fig. 2b) with $C$=-4, -2, 0, …, which resemble energy gaps between Landau levels originating from the Dirac point of the MATBG's band structure. However, we note that these states are also consistent with the selective filling of the $C=\pm1$ bands (see Methods for detailed discussion).

Away from the CNP, the incompressible peaks at ν=+1, +2, and +3 evolve strikingly as the magnetic field increases (Fig. 2c-e): the zero-field peaks rapidly split into multiple states with well-defined Chern numbers, many of which have not been reported so far. The observed incompressible states may be classified according to the parity of $p=C+s$. The even-$p$ states, namely (+1, +1), (0, +2), (+2, +2) and (+1, +3), are generically expected in the presence of $C_2$ symmetry breaking induced by alignment with the hBN substrate, which results in Chern bands with opposite signs at the two valleys. By contrast, the observed incompressible states with $C$=-2 and 0 from ν=+1, $C=\pm1$ from ν=+2, and $C$=0 from ν=+3 each have odd $p$, and cannot be explained by simple flavor polarization in the presence of $C_2T$ symmetry breaking induced by the hBN substrate or by time reversal symmetry breaking[8,12], as both scenarios only result in states with even $p$. The data therefore suggest that further reconstruction of the Chern bands takes place and leads to the observed sequence of Chern numbers. Remarkably, tracking the peaks associated with the incompressible states near $B$=0 T (Fig. 2c-e) reveals that the odd- and even-$p$ states closely compete or coexist at low field, giving rise to the insulating gaps observed on the electron-doped side at zero field.

In contrast to the electron-doped side, the low-field compressibility on the hole-doped side exhibits more conventional characteristics, namely a weak sawtooth pattern and the absence of thermodynamic gaps for -4<ν<0. These observations suggest that the ratio $U/W$, where $U$ is the interaction scale and $W$ is the non-interacting bandwidth, is smaller for holes than for electrons (see Extended Data Fig. 4 and Methods for discussion), consistent with previous reports[13,28,29]. Under such conditions, a sufficiently large perpendicular magnetic field is expected to drive the system into flavor-polarized ChI states due to the strong orbital magnetic moments of the isolated Chern bands[25,33], even if the ground state at zero field is metallic. However, when $B$ reaches 5 T, we begin to detect prominent ChIs with principal Chern numbers $C=-1$ and 0 from $v=-3$, $C=-1$ from $v=-2$, and $C=-2$ and $-1$ from $v=-1$, most of which have odd $p$ and fall outside the sequence predicted by simple $C_2T$ symmetry breaking. Intriguingly, the sequences of ChIs on the electron- and hole-doped sides appear to differ in their energy hierarchies. While on the electron-doped side we observe both even- and odd-$p$ states with comparable strength in inverse compressibility, on the hole-doped side we predominantly detect states with odd $p$, likely due to the fundamental electron-hole asymmetry of the system shown by our transport and compressibility measurements at zero field (see Methods for discussion). Regardless of the asymmetry, the odd-$p$ states on both the electron and hole-doped sides are reproducible and robust against spatial variation (Extended Data Fig. 5), and thus call for a reexamination of the topology and symmetry properties of the system.

The observation of odd-$p$ states suggests that additional symmetry is broken, resulting in reconstruction of the Chern bands. Recent theoretical studies[34,35] have emphasized that, away from the CNP, the charge density modulation in real space significantly modifies the electronic band structure in momentum space. This effect is reflected in the Hartree potential arising from the background electrons for positive $v$ (Fig. 3a), where the inhomogeneous charge distribution raises the energies of the states at the K, K' and M points with respect to those at the Γ point and favors populating electrons around the Γ point. As a result, a new possible ground state at $v=+1$ emerges corresponding to a metal formed by half-filling two of the $C=\pm1$ bands, which competes with the single-valley-polarized ChI. The system can further lower its energy by spontaneously breaking translation symmetry (TS) and folding the Brillouin zone in half[36], which splits the band into two and forms an insulator (Fig. 3d). Because the Berry curvature is highly concentrated at the Γ point (Fig. 3b), the lower band (purple in Fig. 3d) inherits the Chern number of the original band ($C=\pm1$), while the Chern number of the upper band remains zero. On the hole-doped side, the Hartree potential acts with opposite sign, raising the energies of the states at the Γ point and resulting in a lower band with $C=0$ and an upper band with $C=\pm1$. Selectively filling these eight flat bands produces both even- and odd-$p$ states for both electron-and hole-doped sides observed in the experiment (Fig. 3e-j and Extended Data Fig. 6), suggesting that the odd-$p$ states spontaneously break TS.

To further show that the odd-$p$ states are energetically competitive, we have performed calculations within a restricted self-consistent Hartree-Fock approximation (see Supplementary Information Section II). For the odd-$p$ case, we take as trial wavefunctions a family of stripe states similar to those proposed in recent density matrix renormalization group[36,37] and exact diagonalization[38] studies. These states fold the hexagonal moiré Brillouin zone into a rectangle, breaking not only TS but also $C_3$ symmetry. Consistent with previous theoretical studies[36–38], we find that the energy differences between odd- and even-$p$ states at zero magnetic field are very sensitive to the ratio $w_0/w_1$, where $w_0$ is the interlayer tunneling at the AA sites and $w_1$ is that at the AB sites, and can be as small as 0.5 meV per particle at $v=3$ when $w_0/w_1$ is large (Extended Data Fig. 7). These calculations reveal that enhancements of the bandwidth, such as that from the Hartree potential, increase the energy gain associated with forming the TS-breaking states, suggesting an explanation for the dominance of the odd-$p$ states on the hole-doped side of the CNP. We find that the gap to the valence remote band is less than 18 meV, approximately 60% of that to the conduction remote band (Extended Data Fig. 8) and significantly smaller than the interaction energy ~23 meV extracted from recent spectroscopic measurements[19,28]. Because the Hartree correction to the bandwidth becomes non-negligible when the gap separating the flat bands from the remote bands is smaller than the interaction strength, we expect significant enhancement of the bandwidth of the valence flat band, favoring the TS-breaking states on the hole-doped side.

Most of the salient features of our compressibility measurements can be captured within a phenomenological model of the free energy of the system in which the magnetization and energy of each ChI's band are taken as free fit parameters (see Supplementary Information Section III). This toy model highlights the importance of including the effect of the finite magnetization, without which only a single ChI emanating from each $v$ is predicted to occur at a given magnetic field[39]. However, the states with $C<0$ on the electron-doped side are not captured within this phenomenological model. We speculate that both hBN alignment—crucial for obtaining bands with $C<0$ on the electron-doped side within our TS breaking framework—and the cascade of phase transitions may introduce corrections to the free energy of the $C<0$ states beyond the terms considered in our model.

While the device we examine here is likely aligned with the hBN substrate, which influences the exact sequence observed in the experiment, our TS breaking mechanism is also applicable to samples without hBN alignment and will enable the emergence of TS-breaking ChIs whenever interactions are strong and the sample is sufficiently clean and homogenous. More broadly, we anticipate TS breaking with incommensurate periodicity to occur and stabilize incompressible states at fractional moiré band filling, analogous to those recently reported in transition metal dichalcogenide moiré superlattices[40,41], but here stemming from Chern bands. Importantly, our proposed mechanism isolates a key distinctive feature of the physics of MATBG: the interplay between the quantum geometry of wavefunctions—as manifested in the distribution of Berry curvature—and strong electron-electron interactions. Finally, our finding of TS-broken states down to zero magnetic field expands the rich phase diagram of MATBG, and highlights the power of local thermodynamic measurements for providing new insights into the competition between different phases in this strongly-correlated electron system.

## Acknowledgements

We acknowledge discussions with Assaf Hamo and Biao Lian. This work was primarily supported by the U.S. Department of Energy, Basic Energy Sciences Office, Division of Materials Sciences and Engineering under award DE-SC0001819. Fabrication of samples was supported by the U.S. Department of Energy, Basic Energy Sciences Office, Division of Materials Sciences and Engineering under award DE-SC0019300. Help with transport measurements and data analysis were supported by the National Science Foundation (DMR-1809802), and the STC Center for Integrated Quantum Materials (NSF Grant No. DMR-1231319) (Y.C.). P.J-H acknowledges support from the Gordon and Betty Moore Foundation's EPiQS Initiative through Grant GBMF9643. A.T.P. acknowledges support from the Department of Defense through the National Defense Science and Engineering Graduate Fellowship (NDSEG) Program. Y.X. and S.C. acknowledge partial support from the Harvard Quantum Initiative in Science and Engineering. A.T.P., Y.X and A.Y. acknowledge support from the Harvard Quantum Initiative Seed Fund. AV was supported by a Simons Investigator award and by the Simons Collaboration on Ultra-Quantum Matter, which is a grant from the Simons Foundation (651440, AV). EK was supported by a Simons Investigator Fellowship, by NSF-DMR 1411343, and by the German National Academy of Sciences Leopoldina through grant LPDS 2018-02 Leopoldina fellowship. P.R.F. acknowledges support from the National Science Foundation Graduate Research Fellowship under Grant No. DGE 1745303. This research is funded in part by the Gordon and Betty Moore Foundation's EPiQS Initiative, Grant GBMF8683 to D.E.P. K.W. and T.T. acknowledge support from the Elemental Strategy Initiative conducted by the MEXT, Japan, Grant Number JPMXP0112101001 JSP and SKAKENHI Grant Number JP20H00354 and the CREST(JPMJCR15F3), JST. This work was performed, in part, at the Center for Nanoscale Systems (CNS), a member of the National Nanotechnology Infrastructure Network, which is supported by the NSF under award no. ECS-0335765. CNS is part of Harvard University.

## Author Contributions

A.T.P., Y.X., J.M.P., P.J.-H. and A.Y. designed the experiment. A.T.P. and Y.X. performed the scanning SET experiment, the temperature-dependent transport measurements and analyzed the data with input from A.Y. S.H.L., A.T.P. and Y.X. performed the transport measurements in the

# Figure 1

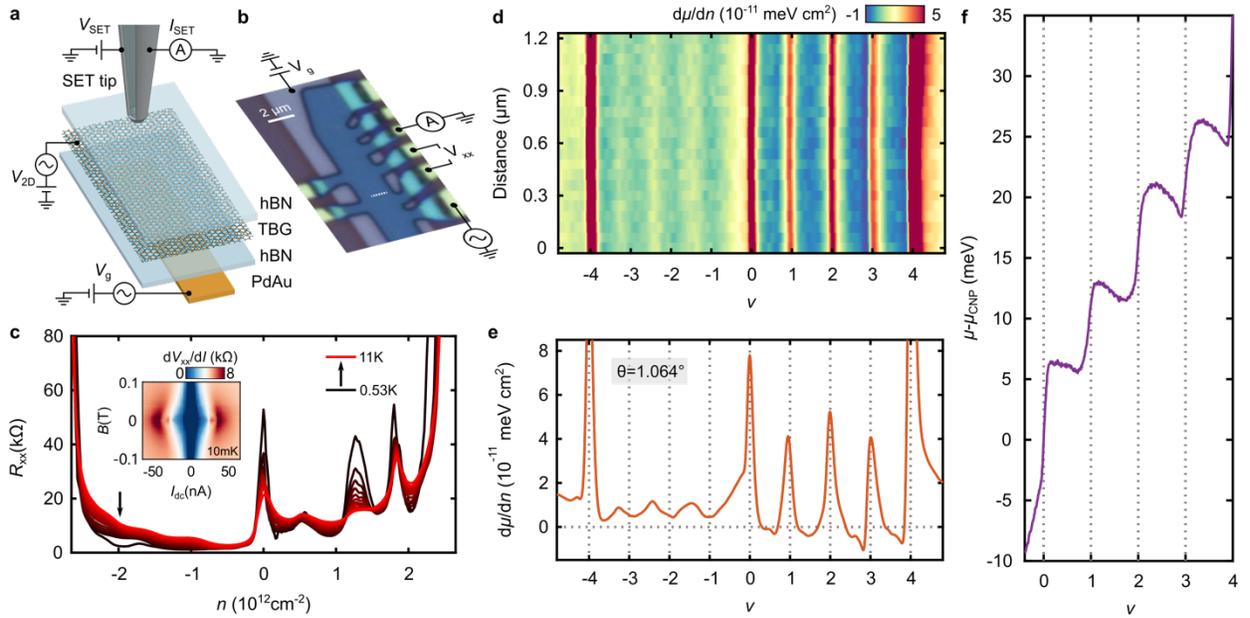

**Fig. 1 | Thermodynamic gaps of correlated insulators in MATBG at zero magnetic field. a,** Schematic of scanning SET measurement setup. **b,** Optical image of device and configuration used for transport measurement. **c,** Four-terminal longitudinal resistance as a function of carrier density measured at different temperatures from 0.53 K (black trace) to 11 K (red trace) using the circuit shown in **b**. Inset: dilution refrigerator ($T \approx 10$ mK) measurement of the differential resistance as a function of applied DC bias current $I_{dc}$ and out-of-plane magnetic field $B$ measured at density $n=-1.97 \times 10^{12}$ cm$^{-2}$ indicated by the arrow in the main plot. **d,** Local inverse compressibility $d\mu/dn$ as a function of moiré band filling factor ν measured along the 1.2 μm white dashed line in **b**. **e,** Spatially averaged $d\mu/dn$ obtained from **d**, showing incompressible peaks aligned precisely at ν=0, 1, 2, 3. **f,** Spatially averaged chemical potential $\mu$ relative to its value at the CNP as a function of moiré band filling factor ν, showing steps at precisely ν=0, 1, 2, 3. At these locations, we find that $\Delta_0$=9 meV, $\Delta_1$=9 meV, $\Delta_2$=10 meV and $\Delta_3$=9 meV (see Methods for gap extraction procedure).

# Figure 2

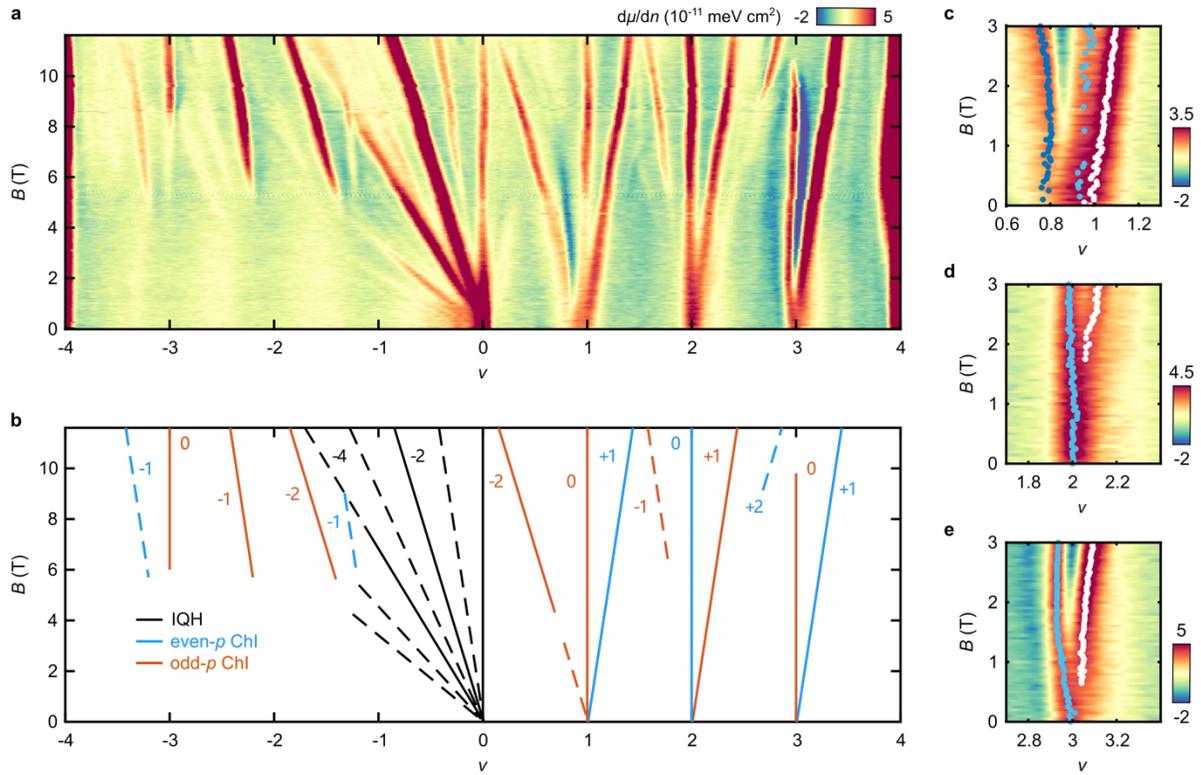

**Fig. 2 | Unconventional sequence of Chern insulators. a,** Local inverse compressibility d$\mu$/d$n$ as a function of moiré band filling factor ν measured between $B$=0 T and 11.6 T. **b,** Schematic Wannier diagram corresponding to the observations in **a**. Strongest features are denoted by solid traces and weaker traces by dashed lines. Integer quantum Hall states (IQH) emanating from the CNP are depicted in black, and ChIs emanating from ν=±1, 2, and 3 are depicted in blue for even-$p$ and orange for odd-$p$. **c-e,** Zoom-in view of **a** around ν=1, 2 and 3. Dark blue, light blue and white circles correspond to incompressible peak positions for $C$=-2, 0 and +1, respectively.

# Figure 3

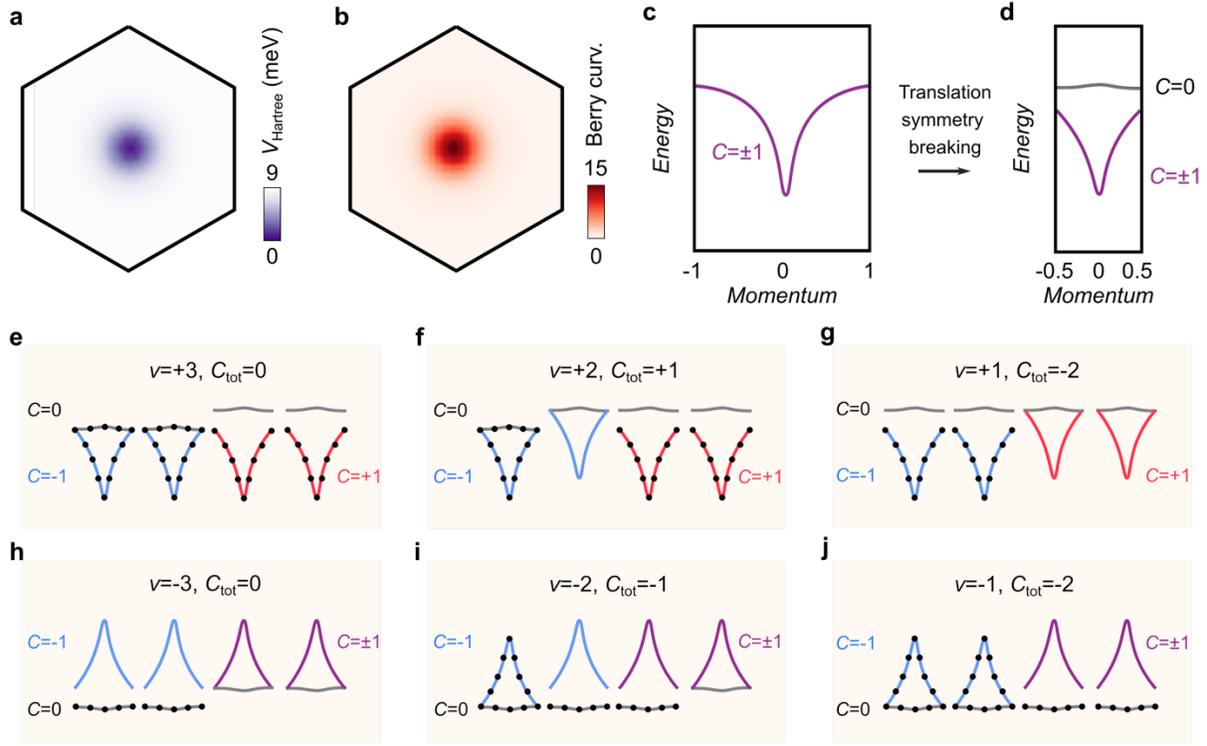

**Fig. 3 | Theoretical model with broken translational symmetry. a, b,** Hartree potential (**a**) and Berry curvature (**b**) for the conduction flat bands within the first moiré Brillouin zone. The center and corners of the hexagon correspond to the Γ and K (K') points, respectively. **c, d,** Sketch of the dispersion of the conduction flat bands and their associated Chern numbers under the influence of the Hartree potential (**c**) and that with spontaneous TS breaking (**d**). **e-j,** Schematic depictions of the band fillings (black circles indicate a filled band) that produce the representative TS-broken (odd-$p$) states with Chern numbers observed experimentally for $\nu = +3$, +2, +1 (**e-g**) and for $\nu = -3, -2, -1$ (**h-j**) (See Extended Data Fig. 5 for the full sequence). We note that only two $C=-1$ bands are necessary to produce the observed sequence on the hole-doped side.